\documentclass{article}\usepackage{natbib,emulateapj} 
\newcommand{\beq}{\begin{equation}}
\newcommand{\eeq}{\end{equation}}
\bibpunct[,]{(}{)}{;}{a}{}{,}
\begin{document}
\title{Particle Heating by Nonlinear Alfv\'enic Turbulence in ADAFs}
\author{Mikhail V. Medvedev\altaffilmark{1} }
\affil{Harvard-Smithsonian Center for Astrophysics,
60 Garden St., Cambridge, MA 02138}
\altaffiltext{1}{Also at the Institute for Nuclear Fusion, RRC ``Kurchatov
Institute'', Moscow 123182, Russia; E-mail: mmedvedev@cfa.harvard.edu;
URL: http://cfa-www.harvard.edu/\~{ }mmedvede/ }

\begin{abstract}
Particle heating in advection-dominated accretion flows (ADAFs) by nonlinear
MHD (Alfv\'enic) turbulence is investigated. Such turbulence with 
highly-fluctuating magnetic fields, $\tilde B\sim B_0$, is believed to be 
naturally produced by the magnetic shearing instability near the nonlinear 
saturation. It is shown that the energy is dissipated in the parallel cascade, 
which occurs due to nonlinear compressibility of high-amplitude turbulence, and
predominantly heats protons, but not electrons.
The conservative limit on the electron--to--proton heating fraction 
is $\delta\lesssim{\rm few}\times10^{-2}$.
\end{abstract}

\section{Introduction}

The Advection-Dominated Accretion Flows (ADAFs) are a class of hot, optically
thin accretion solutions 
\citetext{\citealt{Ichimaru77,Rees-etal82,NY95,Abramowitcz-etal95}; 
see e.g., \citealt{NMQ98} for review}
which describe quite well the spectral characteristics of a number of low 
luminosity accreting black hole systems, e.g., black hole binaries 
\citep{NBMc97,Hameury-etal97} and low luminosity galactic nuclei
\citetext{e.g., Sgr A$^*$: 
\citealt{NYM95,Manmoto-etal97,Mahadevan98,Narayan-etal98}; 
NGC 4258: \citealt{Lasota-etal96,GNB98};
M87 and other ellipticals: \citealt{Reynoldsetal96,Mahadevan97,DMetal98}}.
In ADAFs, the protons and electrons are thermally decoupled
(Coulomb collisions are rare), so that all the 
energy generated by turbulent viscous stresses is stored as thermal energy 
of the protons and ultimately advected beyond the black hole horizon,
while the electrons remain relatively cool and radiate much less energy, hence 
the sub-Eddington luminosities. In such hot accretion flows, however, the
particle mean free path is often comparable to the size of the system, and thus 
collective plasma effects are likely to be significant \citep{Rees-etal82}.

In the absence of collisions, strong MHD turbulence is required for 
the angular momentum transport and energy dissipation in accretion flows.
The way MHD turbulence dissipates in hot accretion flows turns out to be 
crucial for the relevance of ADAF models. Let's denote $P_e$ and $P_p$ the 
amounts of energy that heat the electrons and protons, respectively. 
In traditional ADAF models, the branching ratio parameter $\delta=P_e/P_p$ is 
commonly assumed to be less than or similar to $10^{-2}$. The predicted spectra 
are only weakly sensitive to the exact value of $\delta$ so long as 
$\delta\lesssim10^{-2}$, while for larger values of $\delta$ the changes in 
spectrum characteristics are drastic because the electrons become hot and 
radiate a large fraction of the internal energy of the flow via the synchrotron 
emission so that the flow is no longer advection-dominated. The above
constraint on the value of $\delta$ has been recently relaxed in the 
modified ADAF model with wind outflows \citep{BB98}. Because of the lower 
accretion rates (due to the mass loss) at the inner regions, 
where most of the radiation is produced, the density of the inflowing gas and 
the magnetic field strength decrease (assuming constant $\beta$) and result in
a much lower radiative efficiency of the flow. It was demonstrated by 
\cite{QN98} that the values of $\delta$ as high as $\delta\sim0.3$ are still 
plausible in the advection-dominated flows with strong winds.
The question of turbulent particle heating 
and non-thermal coupling of electrons and protons in hot accretion flows has 
been addressed in several works \citep{Quataert98,QG98,BkL97,BC88,NY95}. 
To proceed further, we need to understand
what type of MHD turbulence is most likely to be realized in ADAFs.

It is commonly believed that MHD turbulence in accretion flows is generated 
by the magneto-rotational instability \citep{BalbusHawley91}, which 
generates the large-scale magnetic field as well as produces random gas 
motions and strong fluctuating magnetic fields in the nonlinear regime. Hence, 
a variety of plasma waves may be generated. Only low-frequency waves may be 
efficiently excited by large scale motions and carry a significant fraction 
of the turbulent energy. \citet{Quataert98} shows that among the low-frequency
MHD waves, fast and slow magnetosonic waves are strongly damped in the
two-temperature  $T_p\sim10^{12}\textrm{ K},\ T_e\sim10^9\textrm{ K}$ 
($T_p$ and $T_e$ are the proton and electron temperatures) ADAF plasma
by collisionless dissipation (known as Landau damping and 
its magnetic analog the transit time damping) while Alfv\'en waves are
still very weakly damped in the incompressible approximation.
As argued by \citet{Quataert98}, the ``impedance mismatch'' will likely inhibit
excitation of heavily damped modes and thus the MHD turbulence in ADAFs will 
be (predominantly) {\em Alfv\'enic}. Additionally, it is natural to believe
that the turbulent fluctuating magnetic fields are of order of the mean, large
scale magnetic field, $\tilde B\sim B_0$, and the Balbus-Hawley instability is 
nonlinearly saturated.

A detailed study particle heating processes by Alfv\'enic turbulence in ADAFs
neglecting the effects of compressibility has been performed by
\citet{Quataert98} and \citet{QG98}. They assumed that the energy which is
injected into the system on large scales comparable to the size of the 
accretion flow, $L\sim R$, will cascade in $k$-space to small scales, 
typically the proton Larmor scale, $l\sim\rho_p={v_{\rm th}}_p/\Omega_p$ 
(with ${v_{\rm th}}$ being the thermal particle
velocity and $\Omega=eB_0/mc$ being the cyclotron frequency) or even smaller,
where it dissipates. Since the Alfv\'en wave cascade is likely anisotropic
\citep{GS95}, the $k$ components are related as 
$k_\|\sim k_\bot^{2/3}L^{-1/3}$. Thus, when the cascade reaches $k_\bot$ such
that $k_\bot\rho_p\sim1$, the wave energy dissipates only via Landau damping 
(Cherenkov resonance) and may preferentially heat either the electrons or the
protons, depending on the value of plasma $\beta$ ($\beta=8\pi p/B_0^2$ is the
ratio of gas to magnetic pressure).\footnotemark
\footnotetext{The plasma physics $\beta$ used in this paper is different 
	from that used in ADAF models and is related to it as 
	$\beta_{\rm adaf}=\beta/(\beta+1)$.}
The cyclotron damping which may be very 
efficient to heat protons is unimportant in this case, because $k_\|\ll 
k_\bot$ and $\omega=k_\|v_A\ll\Omega_p$ (where $v_A^2=B_0^2/4\pi m_pn$ is the 
Alfv\'en speed and $n$ is the particle density), so that the cyclotron resonance
($\omega-k_\|v-\Omega_p=0$) is satisfied for a negligibly small population of
very energetic particles from the tail of the Maxwellian distribution function.
\citet{QG98} also argue that only a fraction of Alfv\'enic energy may be 
dissipated on the scale $k_\bot\sim\rho_p^{-1}$ and the rest of it passes this 
`barrier'. On scales $k_\bot\gg\rho_p^{-1}$ Alfv\'en waves may be converted 
to whistler waves and cascade even further where dissipation on electrons
may be dominant. Because the details of how Alfv\'en waves are converted into
whistler waves on scales $k_\bot\sim\rho_p^{-1}$ are not known, they estimate 
that the transition from the electron dominated heating regime to the proton 
dominated one occurs somewhere inbetween $\beta\sim5$ and $\beta\sim10^2$.

The above analysis is accurate for low-amplitude turbulence, 
$\tilde B/B_0\ll1$, where one can neglect the effect of finite magnetic field
pressure, $\tilde B^2/8\pi$, (i.e., incompressible plasma) 
and use the linear theory. In a high-amplitude turbulence, $\tilde B/B_0\sim1$,
which is likely to be present in accretion flows,
finite magnetic pressure of waves nonlinearly couples Alfv\'enic energy
to ion-acoustic-like (i.e., density) quasi-mode perturbations, which are, in
general, dissipative. In this paper, we show that:
\begin{enumerate}

\item
The Alfv\'enic {\em parallel} cascade towards higher $k_\|$ exists 
in a compressible MHD turbulence as a result of nonlinear wave steepening 
(modulational instability), in addition to the perpendicular 
(incompressible) cascade in $k_\bot$. This parallel cascade is 
nonlinearly dissipative because it proceeds via the excitation of ion-acoustic 
(i.e., compressible) quasi-modes which are damped via the Landau mechanism.
Even for relatively small amplitudes of turbulence, $\tilde B/B_0\gtrsim0.2$, 
and for $\beta\gtrsim3$, much or almost {\em all} magnetic energy of Alfv\'en 
waves dissipates before it reaches the perpendicular `dissipation barrier', 
$k_\bot\rho_p\sim1$. The perpendicular cascade, thus, turns out to be 
energetically unimportant. Dissipation in the nonlinear parallel cascade 
preferentially heats the protons and yields 
$\delta\lesssim{\rm few}\times10^{-2}$.

\item
If $\beta\lesssim3$, the turbulence bifurcates to another `phase' where the
Afv\'enic dissipation in the parallel cascade is weak and other dissipation
mechanisms (at small scales) are required to maintain a steady-state.
Since the parallel cascade proceeds up to 
$k_\|\simeq4(\tilde B/B_0)(\Omega_p/v_A)$, the Alfv\'enic turbulent energy may 
be efficiently dissipated due the cyclotron damping on protons,
if $\omega=k_\|v_A\simeq\Omega_p$, i.e., if $\tilde B/B_0\gtrsim0.3$. Only 
protons may be heated in this regime. For lower amplitudes, the perpendicular 
Goldreich-Sridhar cascade is likely to dominate and energy dissipates, as
discussed by \citet{Quataert98} and \citet{QG98}. The three-wave cascade 
\citep{NB96}, if it dominated the above four-wave cascade, turns out to 
be insignificant for electron energetics.
\end{enumerate}
We also consider other `higher-order' nonlinear effects, such as particle
trapping, and their relevance to particle heating in ADAFs.

The paper is organized as follows. In \S \ref{S:N-KNLS} we discuss the
noisy-KNLS model of strong MHD turbulence. In \S \ref{S:ADAF} we apply the
model to the conditions in ADAFs. We compare the model at hand with other 
competing turbulent processes in \S \ref{S:COMPETE}.
Section \S \ref{S:CONCL} is the conclusion.

\section{The model of nonlinear Alfv\'en wave turbulence \label{S:N-KNLS}}

\subsection{General considerations}

It is known that in an incompressible plasma, $\nabla\cdot{\bf v}=0$, the
Reynolds and magnetic stresses (i.e., the fluid and magnetic nonlinearities) 
in a finite-amplitude Alfv\'en wave cancel each other exactly.
Such a wave behaves as if it is linear. Hence, only the Goldreich-Sridhar
Alfv\'enic perpendicular, $k_\bot$, cascade occurs in a turbulent regime.
The mutual cancellation of Reynolds 
and magnetic stresses in the wave breaks down when plasma is compressible. 
Indeed, the magnetic field pressure, $\tilde B^2/8\pi$, associated with the 
wave field exerts an additional stress onto an ionized gas and change its
local density, $n$. Variations in $n$ affect, in turn, the local wave phase 
speed, $v_A$, and introduce a positive nonlinear feedback into the wave 
dynamics \citep{CohenKulsrud71}. This nonlinear phase speed--amplitude 
coupling in {\em nonlinear Alfv\'en} waves results in the {\em modulational 
instability} which is ultimately responsible for the wave-front steepening, 
parallel, $k_\|$, Alfv\'enic cascade, and formation of collisionless shocks. 
The compressional nonlinearity may also be viewed as an effective parametric 
coupling of a finite-amplitude Alfv\'en wave to ion-acoustic (i.e., density)
wave-like perturbations in a medium.\footnotemark
\footnotetext{
	Such ion-acoustic modes are {\em not} plasma eigenmodes. 
	They are driven by the Alfv\'en wave and, hence, propagate with the 
	Alfv\'en speed, which is, in general, different from the sound speed in
	plasma.}
These ion-acoustic quasi-modes are compressional and have a longitudinal
component of the electric field (with respect to the ambient magnetic field).
Hence, they efficiently dissipate the energy via Landau damping.
A quantitative theory of nonlinear Alfv\'en waves is now well established
\citetext{see a short review by \citealt{KNLSreview} and references therein}.
Note that the dissipation of the finite-amplitude Alfv\'en waves is
intrinsically nonlinear process and leads to a number of unusual properties
of such waves \citep{KNLSreview}.

\subsection{The model \label{S:MODEL}}

There is only one (to our knowledge) model of strong nonlinear Alfv\'en wave 
turbulence in a compressible plasma which self-consistently includes the
effects of wave nonlinearity and particle kinetics, e.g., Landau damping
\citep{MD97}. It is formulated as a stochastic modification of the 
dynamic equation of Alfv\'en waves, --- the kinetic nonlinear Schr\"odinger
equation (KNLS). This model of turbulence, referred to as the ``noisy-KNLS,'' 
is thus structure-based, meaning that it describes the turbulence as a
collection of strongly interacting {\em coherent} structures, such as 
shocks, solitons, cnoidal waves, rotational discontinuities, 
etc., generated by an external random noise source.\footnotemark
\footnotetext{In this respect, the model is analogous to the noisy-Burgers 
	model of hydrodynamic turbulence which describes the turbulence
	as a collection of randomly interacting collisional shocks.}
The noisy-KNLS equation is
\beq
\frac{\partial b}{\partial \tau}+v_A\frac{\partial}{\partial x}
\left(N_1 b\widetilde{|b|^2}
+N_2 b\hat{\cal H}\bigl[\widetilde{|b|^2}\bigr]\right)
+ i\frac{v_A^2}{2\Omega_p}\frac{\partial^2 b}{\partial x^2}=\tilde f ,
\label{noisy-knls}
\eeq
where $\hat{\cal H}$ is the {\em integral operator}, referred to as the Hilbert
operator (or the Hilbert integral transform), acting on the wave field:
\beq
\hat{\cal H}\bigl[\widetilde{|b|^2}\bigr]=\frac{1}{\pi}
\int_{-\infty}^{\infty}\frac{{\cal P}}{(x'-x)}
\widetilde{|b(x')|^2}{\rm d}x' .
\eeq
Here also $b=(\tilde B_y+i\tilde B_z)/B_0$ is the normalized complex wave
magnetic field, $\widetilde{|b|^2}=|b|^2-\langle|b|^2\rangle$ is the
fluctuating component of the magnetic pressure, $\langle\dots\rangle$ means
appropriate averaging over the spatial domain, ${\cal P}$ means principal
value integration, $\tilde f$ is the random (turbulent) noise, and the 
coefficients $N_1$ and $N_2$ are complicated functions of parameters of a 
plasma and are discussed below.

The first nonlinear term in equation (\ref{noisy-knls}) describes the nonlinear 
(ponderomotive) coupling of finite-amplitude Alfv\'en waves to ion-acoustic 
modes. Indeed, the first two terms are similar to a continuity equation for 
the wave magnetic field, $\partial_\tau b+\partial_x(ub)=0$, where the 
``hydrodynamic'' velocity, $u$, is generated by the wave pressure, 
$u\propto v_A\widetilde{|b|^2}$. The process of nonlinear steepening stops 
when it is balanced by linear Alfv\'en wave dispersion which occurs at very 
small scales comparable to proton's Larmor radius, $k_\|\sim\rho_p^{-1}$, 
as represented by the last, linear in $b$, term.
The second nonlinear term in equation (\ref{noisy-knls}) describes Landau
(i.e., collisionless) damping of the induced density perturbations, not the 
magnetic fluctuations. This collective kinetic effect is represented by the
integral Hilbert operator. We emphasize that the 
damping is {\em nonlocal} which can be interpreted as the effect of particle 
``memory'' which is associated with particle transit through the wave packet.
Another interesting and important feature is the absence of any particular
scale beyond which the damping dominates. Indeed, the Fourier representation
of the Hilbert operator $\hat{\cal H}[f]$ acting on a function $f$ is 
$(ik_\|/|k_\||)f_k\equiv i\,{\rm sgn}(k_\|)\,f_k$, i.e., independent of the 
magnitude of $k_\|$. Thus, Landau damping of nonlinear Alfv\'en waves occurs 
at all scales; it is {\em scale invariant}.
We should comment that both nonlinearities result in the wave steepening and 
formation of sharp fronts (shocks) which is, practically, equivalent to the 
parallel cascade of wave energy to smaller scales. Unlike the conventional,
incoherent cascade paradigm, however, this cascade proceeds
through random interactions of coherent wave structures (such as shocks, 
nonlinear waves, etc.). Hence, collective wave--particle interactions,
such as the nonlinear Landau damping and particle trapping also contribute.

We now briefly discuss basic assumptions used in the noisy-KNLS model. First,
equation (\ref{noisy-knls}) is the {\em envelope} equation. That is, Alfv\'en
waves themselves (carrier waves) are assumed to be linear and, thus, 
obey the dispersion $\omega=k_\|v_A$ while the amplitude of these waves may 
vary in space and time and is described by equation (\ref{noisy-knls}). The time
variable $\tau$ is the ``slow'' time of the large-scale dynamics of the
wave envelope, $\tau=(B_0/\tilde B)^2t_A$, where $t_A=\omega^{-1}$ is the
typical ``Alfv\'enic'' time. It has been assumed in derivation that 
$(\tilde B/B_0)^2$ is small but finite. Comparison of the predictions of
this model with observations of nonlinear Alfv\'en waves in the solar wind by 
spacecrafts shows that the KNLS theory is ``robust'' and supports its validity 
even on the margin of applicability, $(\tilde B/B_0)\simeq1$. 

Second, the noisy-KNLS also assumes that waves are propagating in one 
direction, only. The existence of counter-propagating waves allows for 
another type of parametric wave--wave interactions, ---
the decay instability, due to which an Alfv\'en wave may decay into an acoustic
wave and a counter-propagating Alfv\'en wave, --- which greatly complicates any 
analytical treatment of the problem. The decay instability, however, may be
greatly suppressed in hot, two-temperature ADAF conditions. An
acoustic (compressible) wave that is to be generated is heavily damped by
collisionless dissipation. The same argument that the ``impedance mismatch''
inhibits resonant excitation of strongly damped modes suggests here (but does 
not prove, though) that the parametric resonance discussed above will be
inefficient.

Third, the KNLS-based theory considers planar waves, only. That is, no wave
packet modulations are allowed in the plane perpendicular to the direction of
wave propagation; the wave fronts are always flat. The dynamics of waves is, 
thus, purely one-dimensional. This strong limitation of the dynamics and 
evolution of wave structures in the Alfv\'enic turbulence has, probably, 
very little effect (if any) on the processes of particle heating.
Indeed, on large scales, particles move along field lines, only. Thus, 
heating of particles may occur (in Cherenkov resonance) via increase of 
their parallel velocity, only. The perpendicular wave evolution, therefore, 
hardly affects particle energy, unless $k_\bot\sim\rho_p^{-1}$. 
Note, the effective one-dimensionality of the 
problem does not mean that waves may propagate along the ambient magnetic 
field, ${\bf B_0}$, only. It has been shown that waves propagating at angle
are still described by equation (\ref{noisy-knls}) if one formally writes the 
wave field as $b=(\tilde B_y+B_0\sin{\Theta}+i\tilde B_z)/B_0$, where 
$\Theta$ is the angle\footnotemark
\footnotetext{ Strictly speaking, the angle $\Theta$ cannot be pushed close
	to $90^\circ$ because dispersion becomes different for left-hand-
	and right-hand-polarized waves (dispersion on electrons vs. dispersion
	on protons). This sets the limit on the angle 
	$(\pi/2-\Theta)\gg\sqrt{m_e/m_p}$. }
between ${\bf k}$ and ${\bf B_0}$. 

Fourth, the nonlinear coupling coefficients, $N_1$ and $N_2$, presented in
the Appendix are calculated by solving the particle kinetic (Vlasov) equation 
\citep{Spangler89,Spangler90} and agree with those calculated using the guiding 
center formalism \citep{MW88}. Both methods, however, used perturbative
techniques and do not include nonlinear effects such as particle trapping. 
Maxwellian distribution was assumed for bulk particles. \citet{GQ99} showed
that in a strong Goldreich-Sridhar turbulence the time-asymptotic distribution 
function of protons may be different. It is not clear whether this situation
occurs in ADAF because (i) the infall (accretion) time may be shorter than
the velocity diffusion time required to establish a new distribution and 
(ii) the emergent distribution should, in general, be anisotropic (because 
the turbulence is anisotropic) and thus may be unstable against, for instance, 
fire-hose instability which will isotropize and thermalize the particles.
Since Landau damping is a resonant process, it cares about the local slope
of a distribution function, but its overall shape is irrelevant. Therefore, 
using the Maxwellian distribution is reasonable. 

Finally, we do not specify the origin of random forcing for the moment.
In general, involvement of a particular mechanism would force re-ordering
in equation (\ref{noisy-knls}) and changes in $N_1$ and $N_2$. 
This complication is automatically eliminated in the renormalization-group
approach, the key idea of which is to use bare nonlinear couplings to
compute the noise-renormalized (i.e., turbulent) quantities \citep{MD97}.
Going ahead, the reason why the bifurcation point (see next section) seems 
to be determined by the bare $N$'s and not the turbulent ones is simple: 
both nonlinearities are cubic and renormalize in the same way.

\subsection{The structure of Alfv\'enic turbulence}

We are interested in a steady-state turbulence in which the energy input
due to random forcing is balanced by Landau dissipation. Since this 
dissipation is independent of $|k|$, it occurs at {\em all scales}; thus,
no inertial interval exists. An analytical study of the model for arbitrary 
noise is an extremely laborious task. That is why it has been analyzed only 
in the simplest 
case of $\delta$-correlated in space and time, zero-mean, white in $k$ noise
\citep{MD97}. A one-loop renormalization group technique has been utilized.
The noise-dependent renormalized turbulent transport coefficients has been
obtained. Quite importantly, they are mostly determined by properties of the 
noise source on the largest scales (i.e., the system size), --- the so called 
infrared divergences. The large-$k$ tail of the noise spectrum seems to be 
relatively insignificant to control global properties of turbulence.
Thus, the assumption of white noise is likely to be unproblematic 
in the context of turbulence in hot accretion flows, because the spectrum of
turbulence is not known, anyway.

The existence of two distinct {\em phases} (or regimes) of compressible 
Alfv\'enic turbulence constitutes the most important and interesting 
prediction of the theory. The system will {\em bifurcate} (rather than 
smoothly transit) from one phase to another, as plasma parameters smoothly
change. The bifurcation point is determined by the ratio of the 
nonlinearity-to-damping coefficients:
\beq
\left.\frac{|N_1|}{|N_2|}\right|_{\rm bif}\simeq1.3\, .
\label{bif}
\eeq
This result has been obtained rigorously from the analysis of the 
self-consistency of the infrared cutoff scalings of the solutions for turbulent 
transport coefficients. The physical interpretation, however, is very
simple and clear. If $|N_1|/|N_2|<1.3$, the collisionless dissipation always 
dominates the nonlinear parallel energy cascade (i.e., the wave steepening). 
Most of the energy injected on a large scale dissipates {\em before} it reaches 
the dispersion scale, and {\em all} the energy is dissipated {\em via} Landau
damping in the Alfv\'enic cascade. Thus, the steady-state, so-called 
``hydrodynamic'' ($\omega,k\to0$) turbulence with no steep fronts is predicted.
In the opposite case $|N_1|/|N_2|>1.3$, no mathematically rigorous prediction 
can be made, since no solution to renormalized equations exists. Speculatively, 
however, it is clear that the collisionless damping is weak compared to the 
nonlinear steepening. Only a small fraction of the injected energy
dissipates, while most of it cascades along $k_\|$ to the smallest scales set by
the linear dispersion. Since the source continuously injects energy into the 
system, it will be {\em accumulated} at the smallest scales. 
Hence, a  non-steady-state,
small-scale turbulence of Alfv\'enic shocklets is expected. Stationarity may 
be enforced by including other dissipation mechanisms into the model. 
The most efficient one is the cyclotron damping on protons because 
it operates at nearly the same scales 
(i.e., the proton Larmor radius) where the wave energy is accumulated.
Another mechanism which dominates for linear and weakly nonlinear waves 
($\tilde B/B_0\ll1$) is the perpendicular cascade \citep{GS95,NB96}. The energy 
may cascade through $k_\bot\sim\rho_p$ to smaller scales where Alfv\'en waves 
are probably converted into whistler waves. How energy is dissipated in
this case is unclear \citep{QG98}.
The discussion of these issues and quantitative estimates of the rates
of the perpendicular vs. parallel cascade and cyclotron damping are given in 
\S \ref{S:COMPETE}.

\section{Application to hot accretion flows \label{S:ADAF} }

\subsection{Preliminary notes}

In this section, we primarily focus on the ADAF solutions which are the only
known thermally stable, self-consistent models of hot, two-temperature
accretion flows. The self-similar scalings of plasma parameters relevant for
our problem are \citetext{\citealt{NY95}, see also \citealt{Quataert98}}:
\begin{mathletters}
\begin{eqnarray}
T_p&\simeq&2\times10^{12}\frac{\beta}{\beta+1}r^{-1}\textrm{ K}, \\
T_e&\sim&10^9-10^{10}\textrm{ K}, \\
\beta&=&const.,\\
B_0&\simeq&10^9\alpha^{-1/2}(1+\beta)^{-1/2}m^{-1/2}\dot m^{1/2}r^{-5/4}
\textrm{ G}, \\
\frac{\rho_p}{R}&\simeq&6\times10^{-9}\alpha^{1/2}\beta^{-1/2}
m^{-1/2}\dot m^{-1/2}r^{-1/4}, \\
\frac{v_A}{c}&\simeq&0.9(1+\beta)^{-1/2}r^{-1/2}, \\
\frac{v_r}{c}&\simeq&0.37\alpha r^{-1/2}, 
\end{eqnarray}
\label{adaf}
\end{mathletters}
where $\alpha$ is the Shakura-Sunyaev turbulent viscosity parameter, 
$m=M/M_{\sun}$ is the mass of the central object in solar mass units,
$\dot m=\dot M/\dot M_{\rm Edd}$ is the accretion rate in Eddington units
($\dot M_{\rm Edd}=1.39\times10^{18}m\textrm{ g s}^{-1}$), and $r=R/R_S$ is the
local radius in Schwarzschild units ($R_S=2.95\times10^5m\textrm{ cm}$).
Here $B_0$ is the strength of the large-scale magnetic field, $\beta$ is
assumed to be independent of $R$ and takes, likely, sub-equipartition values
($\beta\gtrsim1-10$), $\rho_p/R$ is the proton Larmor scale in terms of
the local scale of the accretion flow, and $v_r/c$ and $v_A/c$ are the
radial inflow velocity and the non-relativistic Alfv\'en speed,\footnotemark\ 
respectively, in units of the speed of light.
\footnotetext{In the relativistic case, the displacement current cannot be
	neglected in Maxwell's equations. The relativistic Alfv\'en 
	velocity then reads $v_A^{\rm rel}=v_A/\sqrt{1+v_A^2/c^2}$. 
	We neglect this relativistic effect in our paper because it may be 
	significant very close to the inner edge of the flow, near the
	last stable orbit, $r\sim3$, only. }
The electron temperature is nearly constant throughout the accretion flow 
and starts to decrease at large radii, $r\gtrsim10^2-10^3$. The efficient
cooling of relativistic electrons prevents higher temperatures at smaller
radii. Thus the proton-to-electron temperature ratio ranges from
$T_p/T_e\sim10^3$ deeply inside the flow at $R\sim\textrm{few}\times R_S$ 
to $T_p/T_e\sim1$ in the outer regions, $R/R_s\gtrsim10^3$.

The MHD turbulence in hot, collisionless accretion flows is likely to
originate as a result of the magneto-rotational instability 
\citep{BalbusHawley91}. In the linear phase, the Balbus-Hawley instability
predominantly amplifies the toroidal component of the large-scale magnetic
field, so that the fluctuations are weak, $\tilde B\ll B_0$. When the
instability reaches nonlinear amplitudes, it becomes large scale, 
$k_{BH}\sim R^{-1}$, so that long wavelength MHD waves are excited and
the MHD turbulence results, as seen from numerical simulations 
\citetext{see, e.g., a review by \citealt{BH98} and references therein}. 
This turbulence
is generated on large scales and, hence, carries most of the gravitational 
potential energy of the accreting gas. Since the ADAF gas/plasma is 
two-temperature, with the protons much hotter than the electrons, all
compressional MHD modes are heavily damped by the collisionless (Landau and
transit time) damping and thus are hard to excite, as suggested by
\citet{Quataert98}. Thus, only noncompressive (Alfv\'en) MHD modes may 
be efficiently excited and reach nonlinear amplitudes. The nonlinear saturation 
of the Halbus-Hawley instability means that the local fluctuating magnetic 
fields in the accretion flow become comparable to the averaged global magnetic
field and prevent its further growth; thus in the absence of 
dissipation\footnotemark\ $\tilde B/B_0\sim1$.
\footnotetext{This condition does not place any constraint on the value of
$B_0$.} 
At such amplitudes, noncompressibility of Alfv\'en waves is no longer a 
good approximation and the nonlinear theory of strong MHD turbulence 
discussed in the previous section should be used instead. In this case,
Landau damping will decrease the amplitude of turbulence to a level when
dissipation balances energy input. Given the noise properties, the 
amplitude of turbulence may be estimated from \citet{MD97}.

\subsection{Dissipation-dominated parallel cascade}

As we discussed in the previous section (\S \ref{S:N-KNLS}), there is a regime
of strong nonlinear Alfv\'enic turbulence in which all energy that is injected 
into a system on large scales dissipates in the cascade and does not reach the
proton Larmor radius scales. The bifurcation point, which is given by 
Eq.\ (\ref{bif}), is a function of $T_p/T_e$ and $\beta$, only 
(via $N_1$ and $N_2$, see Appendix). It is independent of the fluctuation level
$b\propto\tilde B/B_0$ because steepening and damping are both cubic in $b$
and the wave amplitude cancels out. (The rate of dissipation, however, does
depend on the amplitude of turbulence, as discussed below in 
\S \ref{SSS:RATES-PAR}.) Using the expressions for $N_1$ and $N_2$ given in 
Appendix and Eq.\ (\ref{bif}), we now plot the borderline between the two 
regimes, as a function of plasma parameters. 

The $T_p/T_e$-$\beta$-diagram of state of the 
nonlinear Alfv\'enic turbulence is shown in Fig.\ \ref{fig-state}.
The shaded region corresponds to $|N_1/N_2|<1.3$ and the unshaded region --- 
to $|N_1/N_2|>1.3$. The bold-faced part of the boundary curve represents the
range of the temperature ratio relevant for ADAFs:
$T_p/T_e$ decreases from $\sim10^3$ in the inner regions of the accretion 
flow at radii $R\lesssim10R_S$ to $\sim1$ in the outer regions at 
$R\gtrsim10^2-10^3R_S$. As is seen from the figure, the boundary between the
two phases of turbulence is insensitive to the temperature ratio for the
ADAF conditions and set only by plasma $\beta$. 
If $\beta>\beta_{\rm crit}\simeq2.6$, the Alfv\'enic turbulence in an ADAF 
is in the regime in which the parallel cascade is strongly dissipative.
If additionally the parallel cascade is dominant (i.e., faster than any other
cascade/dissipation process), then it dissipates all injected energy.
Alternatively, for $\beta<\beta_{\rm crit}\simeq2.6$, Alfv\'en wave energy 
does not dissipate in the parallel cascade (even if it is dynamically dominant) 
and reaches $k_\|\sim\rho_p^{-1}$ where it may be dissipated by other 
mechanisms, or cascade beyond this point to even smaller scales. 

\subsection{Electron vs. proton heating in the parallel cascade
\label{SS:EL-HEAT} }

We can also investigate the electron-to-proton heating ratio, $\delta$, using
the equations for $N_1$ and $N_2$ as follows. The electron and proton
contributions to the wave dissipation are proportional to the imaginary part of
the plasma dispersion function, Eq.\ (\ref{plasma-disp}). The factor in
Eq.\ (\ref{plasma-disp}) with integral over the distribution function
has been explicitly incorporated into the wave equation, 
Eq.\ (\ref{noisy-knls}), as the Hilbert operator. The numerical factor
was left in the coefficients, Eqs.\ (\ref{f's}) -- (\ref{g's}), and is 
proportional to $e^{-X^2_e}$ and $e^{-X^2_p}$ for the electrons and protons,
respectively. By vanishing by hands all those terms which are multiplied by 
$e^{-X^2_e}$, we may easily determine how much the electrons contribute to the 
overall dissipation of waves. Calculaing the ratio $(1-N_2^{*})/N_2^{*}=\delta$,
where $N_2^{*}$ is the nonlinear dissipation coupling coefficient without the 
contribution of the electrons to dissipation, we obtain the heating ratio.
Note, that in $N_2^{*}$ the electrons still contribute to the linear and
nonlinear dispersion of waves, as represented by the real part of the plasma
distribution function. 

A contour plot showing the levels of constant heating ratio, 
$\delta=.005,\, .01,\, \dots,\, .035$, is superimposed onto the Alfv\'enic
dominated dissipation region in Fig.\ \ref{fig-state}. Fig. \ref{fig-heat}
also depicts the heating ratio as a function of the temperature ratio 
$T_p/T_e$ for two values of $\beta$,\ $\beta=3$ and $40$.
Clearly, $\delta$ depends on $\beta$ only weakly and ranges with
temperature from $\sim.025$ to $\sim.005$ for the typical ADAF conditions.
Since, the spectral ADAF models weakly depend on $\delta$ so long as
$\delta\lesssim\textrm{few}\times10^{-2}$, we set a conservative upper limit
on the electron heating in the dissipative {\em parallel} cascade of nonlinear
Alfv\'enic turbulence:
\beq
\delta_\|\equiv\left.\frac{P_e}{P_p}\right|_{\|\,{\rm casc}}\lesssim.025 , 
\quad\textrm{ if }\ \beta>\beta_{\rm crit}\simeq2.6 .
\label{delta-limit}
\eeq

\section{Other competing processes \label{S:COMPETE} }

In the previous section (\S \ref{SS:EL-HEAT}) we demonstrated that the 
energy which heats the electrons in the compressible Alfv\'enic dissipative 
parallel cascade in typical ADAF conditions constitues a few percents of the 
total energy which goes into the protons. If this parallel cascade would be 
the only one which operates, then equation (\ref{delta-limit}) gives a solid 
prediction. In fact, there are other competing processes which transfer 
Alfv\'enic energy in $k$-space and thus may affect the energetics. 
We consider them in order.

In addition to the compressible parallel cascade, there is an incompressible 
perpendicular cascade due to direct interactions of linear Alfv\'en wave 
packets. Recently two different theories of this process have been proposed.
\citet{GS95} (hereafter GS) showed that 4-wave interactions lead to a 
critically balanced anisotropic cascade in which $k_\|$ and $k_\bot$ are 
uniquely related. A nice paper by \citet{NB96} (hereafter NB) demonstrates 
that 3-wave interactions of $k_\|=0$ perturbations are not empty and result 
in a more conventional Iroshnikov-Kraichnan-type cascade. Although there are 
some indications from both theoretical arguments and numerical simulations
\citep{GS97,Vishniac00} that the GS cascade dominates over the NB cascade, 
the debate is still not completely resolved. For this reason, we consider both
possibilities below. Briefly, we obtained that the GS cascade puts some 
limitations on our predictions of \S \ref{SS:EL-HEAT}. The NB cascade, 
however, does not alter our conclusions about the electron vs proton heating,
despite that the total wave energy in the perpendicular direction may be
substantial.

\subsection{Parallel vs. GS cascade \label{SSS:RATES-PAR} }

The critically balanced GS cascade \citep{GS95} is an anisotropic, 
(almost) perpendicular cascade in which $k_\|\ll k_\bot\propto k_\|^{3/2}$.
This cascade involves linear Alfv\'en waves and, thus, is independent of the
amplitude of the turbulence. No damping occurs in this process. The MHD 
(Alfv\'en) energy is transferred from the scales where it is injected, 
$k_{\bot i}\sim R^{-1}$, to the scales where it is dissipated or converted into 
other waves, $k_{\bot f}\sim\rho_p^{-1}$. We have to compare it with the
competing compressible parallel cascade during which much of injected energy 
is dissipated. The rate of
the energy transfer depends on the turbulence level via the compressibility.
Since these two cascade processes are independent (at least in the KNLS
approximation), we may compare relative contribution of each of them to 
the electron heating as a function of the amplitude of MHD turbulence in ADAFs.

The characteristic rate (i.e., the inverse $e$-folding time) of the 
GS cascade is the Alfv\'enic frequency, 
$\gamma_{\rm GS}\sim\omega_A$. The characteristic dissipation rate in the
parallel cascade and the nonlinear cascade rate itself are comparable and 
estimated from Eq.\ (\ref{noisy-knls}) to be 
$\gamma_{\rm NL}\sim\omega_A|N_2|\widetilde{|b|^2}$. Note, although the rate
of the GS cascade, $\gamma_{\rm GS}\propto k_\|$, increases
with $k_\|$, so does the energy dissipation rate. Thus, the total energy 
dissipated in the cascade is determined by the width of the ``inertial range,''
$R^{-1}\lesssim k\lesssim\rho_p^{-1}$. The energy dissipation rate is, by
definition, 
\beq
\frac{d\epsilon}{dt}=\frac{d\epsilon}{dk_\|}\frac{dk_\|}{dt}
=-\gamma_{\rm NL}\,\epsilon .
\eeq
Using the definition for the total energy cascade rate, 
$dk_\|/dt=(\gamma_{\rm GS}+\gamma_{\rm NL})k_\|$, and integrating the
resultant equation, we obtain
\beq
\frac{\ln\epsilon/\epsilon_0}{\ln k_{\|m}/k_{\|0}}
=-\frac{|N_2|\widetilde{|b|^2}}{|N_2|\widetilde{|b|^2}+1} ,
\eeq
where $\epsilon_0$ and $k_{\|0}$ are initial values. The largest scale is 
set by the size of the accretion flow, $k_{\|0}\simeq R^{-1}$. The smallest 
scale is estimated from the scaling, $k_\|\simeq k_\bot^{2/3}R^{-1/3}$, of the 
GS cascade with $k_{\bot m}\sim\rho_p^{-1}$.\footnote{
	Note, here we set a lower limit on $k_{\|m}$. Since both
	cascades proceed independently, $k_{\|m}$ may only be larger 
	(i.e., closer to $\rho_p^{-1}$) due to the parallel cascade.} 
Thus, $k_{\|m}/k_{\|0}\simeq (R/\rho_p)^{2/3}$. 
\citet{QG98} argue that much of the energy dissipated at
$k_\bot\sim\rho_p^{-1}$ may heat electrons, depending on the value of $\beta$ 
and numerical constants estimated from numerical simulations which are not 
very well constrained. We assume here the {\em worst} possible case in which 
{\em all the wave energy} reaching the Larmor scale heats electrons.

For the electron heating to dominate by the compressible cascade, the amount of
energy that reaches the Larmor scale must be smaller than the fraction of 
energy that is dissipated on electrons in the compressible cascade: 
\beq
\epsilon(k_{\|m})/\epsilon_0 < \delta_\|,
\eeq
where $\delta$ is given by Eq.\ (\ref{delta-limit}). Combining all
together, this inequality may now be re-written as follows:
\beq
\widetilde{|b|^2}>-\frac{1}{|N_2|}\,\frac{\frac{3}{2}\ln\delta_\|/\ln(R/\rho_p)}
{1+\frac{3}{2}\ln\delta_\|/\ln(R/\rho_p)} .
\eeq
The nonlinear coupling to dissipation, $N_2$, depends on plasma parameters.
The dependence of $|N_2|$ vs. $\beta$ for two values of the
proton-to-electron temperature ratio: $T_p/T_e=1,\ 10^3$
is shown in Fig.\ \ref{fig-rates}. Clearly, it weakly
depends on $T_p/T_e$ and is roughly $\propto\sqrt\beta$. To estimate the lower
bound on $|b|$, we take the {\em lowest} value of $|N_2|\sim1$ for
$\beta\sim3$, i.e., near the bifurcation value. The normalized amplitude may
be written in a general case as follows (see \S \ref{S:MODEL}):
\begin{eqnarray}
\widetilde{|b|^2}&=&(\tilde B + B_0\,k_\bot/k)^2/B_0^2-
\langle(\tilde B + B_0\,k_\bot/k)^2/B_0^2\rangle \nonumber\\
&=&(\tilde B/B_0)^2-\langle(\tilde B/B_0)^2\rangle
+2(\tilde B/B_0)k_\bot/k \nonumber\\
&\sim&2(\tilde B/B_0)
\end{eqnarray}
for $\tilde B\lesssim B_0$ and an oblique and nearly perpendicular 
angle of propagation,
$k_\bot\sim k$. We take [see Eqs.\ (\ref{delta-limit}) and (\ref{adaf})]
$\delta\lesssim2.5\times10^{-2}$ and $\rho_p/R\gtrsim6\times10^{-9}$. Then the
conservative limit on the fluctuation amplitude of MHD turbulence at which the
electron heating is dominated by the compressible cascade is readily estimated
as
\beq
\frac{\tilde B}{B_0}\gtrsim0.2\, .
\eeq
Thus, for $\tilde B/B_0\sim0.2$, the electron heating due to the parallel,
$\delta_\|$, and perpendicular, $\delta_\bot$, cascades are of comparable
value and the total electron heating is $\delta=\delta_\|+\delta_\bot\sim0.05$. 
Hence, for the levels of turbulence which are believed to exist in ADAFs,
$\tilde B/B_0\sim1$, the heating of electrons is dominated by the compressible
cascade by at least an order of magnitude and yields $\delta\lesssim0.025$,
provided $\beta>\beta_{\rm crit}\simeq2.6$.

\subsection{Parallel vs. NB cascade \label{SSS:NB} }

In a similar way we consider competition of the NB and parallel cascades.
The NB cascade \citep{NB96} proceeds in $k_\bot$ only, hence it is totally 
independent of $k_\|$. There is no dissipation during the NB cascade. 
The Alfv\'en wave energy may be damped only when it reaches the 
$k_\bot\sim\rho_p^{-1}$
scale and, as discussed in \S \ref{SSS:RATES-PAR}, will predominantly heat 
the electrons. To estimate the energy lost via the NB cascade, it is sufficient 
to compare the rate, $\gamma_{\rm NB}(\rho_p^{-1})$, of the NB cascade at 
$k_\bot\sim\rho_p^{-1}$ to the maximum damping rate in the $k_\|$ cascade,
$\gamma_{\rm NL,max}$. The latter is $\gamma_{\rm NL,max}\sim
\omega_{A,max}|N_2|\widetilde{|b|^2}\sim v_A\rho_p^{-1}|N_2|\widetilde{|b|^2}$.
The former is estimated from \citet{NB96} as follows. The energy transfer rate 
is 
\beq
\dot\epsilon\sim v^2/(N\tau)\sim\textrm{ constant},
\eeq
where $v^2\sim E_kk$ is the energy per unit mass, $N\sim(\delta v/v)$ is the
number of interactions, $\tau\sim(\kappa_\|v_A)^{-1}$ is the interaction time,
$\delta v$ is the perturbation, and $\kappa_\|$is a longitudinal scale of 
interacting Alfv\'en wave packets. Therefore,
\begin{eqnarray}
\gamma_{\rm NB}&\sim&(N\tau)^{-1}\sim\dot\epsilon/v^2\sim\dot\epsilon(E_kk)^{-1}
\nonumber\\
&\sim&(E_{k0}k_0)^{-1}\left(\frac{k}{k_0}\right)^{1/2}\sim
\left(\frac{\delta v_0}{v_0}\right)^2\kappa_{\|0}v_A,
\end{eqnarray}
where we used the NB scaling $E_k\propto k^{-3/2}$ and the subscript ``0'' 
denotes quantities at the outer scale of the turbulence. Assuming strong 
turbulence, $\delta v_0\sim v_0$, we obtain an upper limit on $\gamma_{\rm NB}$.
Assuming also that on the outer scale $k_0\sim\kappa_0\sim\kappa_{\|0}\sim
\kappa_{\bot0}\sim R^{-1}$, we have
\beq
\gamma_{\rm NB}(\rho_p^{-1})\sim 
\left.v_A(\kappa_0 k)^{1/2}\right|_{k\sim\rho_p^{-1}}\sim
v_A(R\rho_p)^{-1/2}.
\eeq
Thus, the ratio is
\beq
\frac{\gamma_{\rm NB}(\rho_p^{-1})}{\gamma_{\rm NL,max}}\sim
\left(|N_2|\widetilde{|b|^2}\right)^{-1}\left(\frac{\rho_p}{R}\right)^{1/2}
\sim10^{-4} .
\eeq 
Thus, even if all the energy cascaded in $k_\bot$ heats the electrons,
it constitutes about $10^{-4}$ of the total energy dissipated on the protons
(unless $\tilde B/B_0\ll1$), i.e., is completely negligible compared to 
equation (\ref{delta-limit}). Note that the above result means only that
the NB cascade is ``slower'' than the parallel cascade, while the total 
wave energy in $k_\bot$ may be non-negligible.

\subsection{Proton-cyclotron damping vs. perpendicular cascade 
\label{SSS:RATES-CYC} }

The nonlinear interaction of finite-amplitude wave-packets results in 
generation of high-$k$ harmonics, as discussed in \S \ref{S:MODEL}. Such a 
cascade transfers wave energy along the local magnetic field to high $k_\|$, 
where the cyclotron resonance, $\omega-k_\|v-\Omega_p=0$, may be satisfied and 
the cyclotron damping on (bulk) protons becomes very efficient. This effect was
completely missing from previous studies because both the GS and NB cascades
are (nearly) perpendicular; $k_\|$ is always small, so that only a very
small number of particles from the tail of a particle distribution function
are in the resonance. The proton-cyclotron damping is known to heat protons
only, no energy goes to electrons, since $\Omega_p\ll\Omega_e$ and electrons 
are off resonance. It is, however, difficult to rigorously estimate the 
$\widetilde{|b|^2}$, because for the cyclotron damping to dominate, it must be
faster than the typical GS or NB cascading time, i.e.,
$\gamma_c\gtrsim\omega_A$. Such a situation may occur only when
$\omega_A\simeq\Omega_p$, so that (i) Alfv\'en waves are heavily damped and
(ii) they may convert to proton cyclotron waves, which greatly complicates a
rigorous analysis. 

One can, however, roughly estimate the fluctuation level at which the 
proton-cyclotron damping is so strong that most of the wave energy dissipates 
on protons and does not convert into whistlers. The parallel cascade 
proceeds until it hits a scale where the nonlinear steepening is
balanced by the wave dispersion [last term in Eq.\ (\ref{noisy-knls})]. Thus,
the maximum parallel wave-number to which wave energy cascades in the
compressible cascade is readily estimated from Eq.\ (\ref{noisy-knls}) to be
${k_\|}_{\rm max}\sim\widetilde{|b|^2}(2\Omega_p/v_A|N_{1,2}|)$, 
where $|N_{1,2}|$ is the maximum of $|N_1|$ and $|N_2|$. The proton-cyclotron 
damping is strong when $\omega_A=k_\|v_A\simeq\Omega_p$. Thus, if 
${k_\|}_{\rm max}\gtrsim\Omega_p/v_A$, the MHD turbulence heats protons only.
This occurs when
\beq
\frac{\tilde B}{B_0}\gtrsim\frac{1}{4|N_{1,2}|} ,
\eeq
where we again took $\widetilde{|b|^2}\simeq2(\tilde B/B_0)$. Here 
$|N_{1,2}|\sim1$ for $\beta\sim3$ and increases with $\beta$ and 
weakly with $T_p/T_e$ as shown in Fig.\ \ref{fig-rates}. For a range of
$\beta$'s, $1\lesssim\beta\lesssim2.6$, i.e., above the bifurcation threshold
(dissipation in the compressible cascade is weak), but below equipartition,
the conservative condition on $\tilde B$ reads:
\beq
\frac{\tilde B}{B_0}\gtrsim0.3\, .
\eeq

\subsection{Is nonlinear trapping important?}

The parallel (coherent) cascade of the wave energy occurs due to the formation 
and nonlinear evolution of coherent wave structures, such as shocks, Alfv\'enic
discontinuities, solitary wave-packets, etc., as discussed in 
\S \ref{S:MODEL}. The resonant particles, i.e., those which take the energy 
from waves and result in damping, turn out to be trapped in the wave potential 
created by the wave magnetic field. Those resonant particles which move a little
slower than the wave will be reflected from the rear part of the wave potential,
gain energy, and start to move faster than the wave. The particles moving
faster than the wave will be reflected from the front part of the wave
potential and, correspondingly, give energy to the wave and decelerate. Upon a
half bounce time, $\tau_{\rm b}/2$, the particles which have been reflected 
from the rear of the potential cross though it and reach its front part, 
where they are reflected back and give their energy to the wave. The slow 
particles do the opposite at the rear part. Such a process may repeat over 
and over again, provided the wave amplitude is maintained constant by an
external source (e.g., the Balbus-Hawley instability in ADAFs),
to prevent particles from escape. Clearly, the number of slow ($v<v_A$) 
and fast ($v>v_A$) particles will oscillate with time, so that the wave
damping rate will change in time too and may become negative 
(i.e., the wave amplitude will grow) during some periods of time. 
The bounce period of a particle depends on its energy, because a wave trapping 
potential is, in general, anharmonic. Due to the difference in the bounce
periods, groups of particles with different energies gain a phase shift which
grows with time. Upon many cycles (bounces), the particles near the resonance,
$v\sim v_A$, become completely randomized in phase and form an equilibrium,
{\em plateau} distribution. At this moment, the wave collisionless dissipation
quenches.\footnote{This process does not affect the cyclotron damping on
protons, which operates at the cyclotron, and not Cherenkov, resonance.}
The process described above is referred to as the nonlinear 
Landau damping \citetext{see \citealt{Metal98} for more details}.
Numerical simulations of the process show that it takes
$\sim10^2-10^3\tau_{\rm b}$ or even longer for particles to phase mix and 
quench collisionless damping. We note here that the nonlinear Landau damping is
significant for large amplitude waves $\tilde B/B_0\sim1$, for which the
number of trapped particles may be large.

We now estimate whether particle trapping is important for the physics of
ADAFs and, in particular, whether the Alfv\'en wave damping during the parallel 
(compressible) cascade in ADAFs quenches. First,
the kinetic energy of a particle in a potential is of order its potential
energy, so that the typical velocity of a particle of mass $m_j$
($j=e,p$) is $v_j\sim(\tilde B^2/8\pi m_j n)^{1/2}\sim
{v_A}_j(\tilde B/B_0)$. The typical scale of the potential, $\lambda$, 
is set by the scale at which the energy is pumped into the system, i.e., 
the scale of the Balbus-Hawley instability, $\lambda\sim k_{BH}^{-1}\sim R$.
Thus the characteristic oscillation period of a particle (the bounce time) 
at a given radial position in the accretion flow is
$\tau_{\rm b}\sim\lambda/v_j$, which yields
\beq
{\tau_{\rm b}}_p\simeq R/(v_A\tilde B/B_0), \qquad
{\tau_{\rm b}}_e=(m_e/m_p)^{1/2}\,{\tau_{\rm b}}_p 
\eeq
for protons and electrons, respectively.
Second, accreting gas flows into the central object and, thus, continuously
replenished with a fresh material on the outer edge of the ADAF. 
The typical infall time at a given radius $R$ is
\beq
\tau_{\rm r}\simeq R/v_r ,
\eeq
where $v_r$ is the radial velocity of a gas given by Eqs.\ (\ref{adaf}).
Estimating the number of bounces of a particle of species $j$ in an 
Alfv\'en wave at a radial position $R$ as 
${\cal N}_j(R)\sim\tau_{\rm r}/{\tau_{\rm b}}_j$, we obtain
\beq
{\cal N}_p\simeq0.4\alpha(1+\beta)^{1/2}(\tilde B/B_0), \qquad
{\cal N}_e\simeq43{\cal N}_p .
\eeq
Note that ${\cal N}$ is independent of radius. The above equations show that
the protons in the accretion flow are always in the linear regime of damping
(no trapping, ${\cal N}_p\lesssim1$). The electrons may experience 
up to a few tens of bounces
within the infall time, which is probably not enough to quench damping on
electrons at all, but still may lower its value a little. We thus conclude
that the damping process in the compressible parallel cascade is hardly
affected by nonlinear particle trapping and the value of 
$\delta\sim{\rm few}\times10^{-2}$ represents a conservative upper 
limit on the the electron to proton heating ratio in hot, advection-dominated 
accretion flows with strong magnetic field turbulence, $\tilde B\sim B_0$.

\section{Conclusions \label{S:CONCL} }

The particle heating is one of the key issues for the advection-dominated 
accretion flow models. All the models assume that protons receive most of the
energy released due to viscous (turbulent) heating of the accretion gas, while
electrons remain relatively cold; hence the low luminosities of ADAFs.
In the collisionless plasma of the ADAFs, particle heating is determined
solely by excitation and dissipation of plasma collective motions (waves). 
The instability of Balbus \& Hawley which is believed to operate and produce
magnetic fields in such differentially rotating flows may naturally result 
in strong MHD turbulence with highly fluctuating magnetic fields, 
$\tilde B\sim B_0$, during the nonlinear stages of its evolution. A detailed 
analysis of particle heating by such turbulence is presented in this paper.

We show that in the most natural case of the ADAF parameters, 
$\beta\gtrsim3$ and $\tilde B/B_0\gtrsim0.2$, dissipation of the magnetic 
energy occurs predominantly in the parallel, compressible cascade (the effect 
absent for linear, low-amplitude Alfv\'en waves). Most of the dissipated energy 
heats protons, while electrons receive only $\delta\sim{\rm few}\%$ of the 
energy, which in agreement with the empirical values inferred from the spectral 
fits for various, low-luminosity accreting systems.
If, alternatively, the magnetic field in ADAFs is close to equipartition,
$\beta\lesssim3$, the energy of MHD turbulence is dissipated only on protons
via the cyclotron damping mechanism, provided the amplitude of the turbulence
is somewhat higher, $\tilde B/B_0\gtrsim{\rm few}\times10^{-1}$. The electron
heating is negligible in this case. At lower amplitudes, however, the
nonlinear effects are weak and the results of the linear analysis for 
the dissipation of the linear Alfv\'en waves, addressed by other authors, hold.

\acknowledgements
The author is grateful to Ramesh Narayan for his interest in this work and 
helpful discussions, to the referee Amitava Bhattacharjee for many insightful
comments and suggestions which helped to greatly improve the manuscript
and stimulated further work, and to Eliot Quataert for discussions. 
This work was supported by NASA grant NAG~5-2837 and NSF grant PHY~9507695.

\begin{appendix}
\section{Exact formulae for the nonlinear coefficients $N_1$ and $N_2$}

The nonlinear coefficients in Eq.\ (\ref{noisy-knls}) has been calculated from
the full kinetic treatment by solving the Vlasov equation for arbitrary wave
profile and determining the wave-induced plasma density perturbation
\citep{Spangler89,Spangler90}. They are
\beq
N_1=M_1-m_1, \qquad N_2=M_2-m_2 ,
\eeq
where $M_1$ and $M_2$ are the isotropic pressure contributions and
$m_1$ and $m_2$ are the contributions due to pressure anisotropy
(i.e., the temperatures and, hence, pressures may be different along the
ambient magnetic field and perpendicular to it, $p_\|\not=p_\bot$):
\begin{eqnarray}
M_1&=&-\frac{1}{4}
\left(f_7+\frac{f_5\left(f_1f_2-\pi f_3f_4\right)
+\pi f_6\left(f_3f_2+f_1f_4\right)}{\left(f_2^2+\pi f_4^2\right)}\right), \\
M_2&=&-\frac{\sqrt{\pi}}{4}
\left(f_6+\frac{f_6\left(f_1f_2-\pi f_3f_4\right)
-f_5\left(f_3f_2+f_1f_4\right)}{\left(f_2^2+\pi f_4^2\right)}\right), \\
m_1&=&-\frac{1}{4X_p^2}
\left(g_1+\frac{g_4\left(f_1f_2-\pi f_3f_4\right)
+\pi g_3\left(f_3f_2+f_1f_4\right)}{2\left(f_2^2+\pi f_4^2\right)}\right), \\
m_2&=&-\frac{\sqrt{\pi}}{4X_p^2}
\left(g_2+\frac{g_3\left(f_1f_2-\pi f_3f_4\right)
-g_4\left(f_3f_2+f_1f_4\right)}{2\left(f_2^2+\pi f_4^2\right)}\right),
\end{eqnarray}
where $\sqrt{\pi}$ appears in numerators of $M_2$ and $m_2$ because the
Hilbert operator carries an extra ($1/\pi$) and additional (1/2) in all four
coefficients absorbs it from the wave equation, as compared to 
\citet{Spangler89,Spangler90}.
The coefficients $f_1,\ f_2, \dots,\ f_7, g_1, \dots,\ g_4$ are defined as
\begin{eqnarray}
f_1&=&-X_pZ_R(X_p)+X_eZ_R(X_e), \label{f's} \\
f_2&=&[1+X_pZ_R(X_p)]+T_r[1+X_eZ_R(X_e)], \\
f_3&=&X_pe^{-X_p^2}-X_ee^{-X_e^2}, \\
f_4&=&X_pe^{-X_p^2}+T_rX_ee^{-X_e^2}, \\
f_5&=&1+X_pZ_R(X_p), \\
f_6&=&X_pe^{-X_p^2}, \\
f_7&=&X_pZ_R(X_p), 
\end{eqnarray}
\begin{eqnarray}
g_1&=&[X_p^2+X_p^3Z_R(X_p)-X_pZ_R(X_p)]
+(1/T_r)[X_e^2+X_e^3Z_R(X_e)-X_eZ_R(X_e)], \\
g_2&=&(X_p^3-X_p)e^{-X_p^2}+(1/T_r)(X_e^3-X_e)e^{-X_e^2}, \\
g_3&=&(2X_p^3-X_p)e^{-X_p^2}-(2X_e^3-X_e)e^{-X_e^2}, \\
g_4&=&[2X_p^2+2X_p^3Z_R(X_p)-X_pZ_R(X_p)]-[2X_e^2+2X_e^3Z_R(X_e)-X_eZ_R(X_e)].
\label{g's}
\end{eqnarray}
Here $Z_R(X)$ is the real part of the plasma dispersion function 
for argument $X$:
\beq
Z(X)=\frac{1}{\sqrt{\pi}}
\int_{-\infty}^\infty\frac{\exp(-\xi^2)}{\xi-X}\;{\rm d}\xi
=2i\,e^{-X^2}\int_{-\infty}^{iX}e^{-\xi^2}{\rm d}\xi .
\label{plasma-disp}
\eeq
The imaginary part of $Z(X)$ appears explicitly in Eq.\ (\ref{noisy-knls}) 
as the Hilbert integral operator. Physically, $X=X_R+iX_I$ is the ratio of
wave phase velocity to thermal velocity. For Alfv\'en waves, $X_p$ and $X_e$ 
become:
\beq
X_p=\left(1+\frac{T_e}{T_p}\right)^{1/2}\frac{1}{\sqrt\beta}, \qquad
X_e=\left(\frac{T_p}{T_e}+1\right)^{1/2}\sqrt{\frac{m_e}{m_p}}\;
\frac{1}{\sqrt\beta}
\eeq
At last, $T_r=T_p/T_e$ is the proton-to-electron temperature ratio and
$m_e/m_p\simeq1/1836$ is the electron-to-proton mass ratio.
\end{appendix}

\figcaption[adaf-state-elheat.eps]{A $T_p/T_e$-$\beta$-diagram of state of
nonlinear Alfv\'enic turbulence. The shaded region corresponds to the 
compressional cascade dominated regime (phase) of turbulence. The contour lines 
represent the electron-to-proton heating ratio, $\delta$, in MHD turbulence in 
ADAFs with $\tilde B/B_0\gtrsim0.1$. The unshaded region corresponds to the
small-scale dissipation dominated phase, where the Goldreich-Sridhar cascade
is dominant. The boldfaced part of the dividing (bifurcation) line 
corresponds to the range of the $T_p/T_e$ parameter typical of ADAFs.
\label{fig-state} }
\figcaption[adaf-elheat.eps]{The electron-to-proton heating ratio vs. $T_p/T_e$ 
for $\beta=3$ and $\beta=40$. \label{fig-heat} }
\figcaption[adaf-rates.eps]{The coefficients of nonlinear damping, $|N_2|$ 
(solid curves), and nonlinear steepening, $N_1$ (dashed curves), vs. $\beta$ 
for two values of the temperature ratio,\ $T_p/T_e=10^3$ (boldface curves) 
and $T_p/T_e=1$ (thin curves). \label{fig-rates} }

\vfill
\plottwo{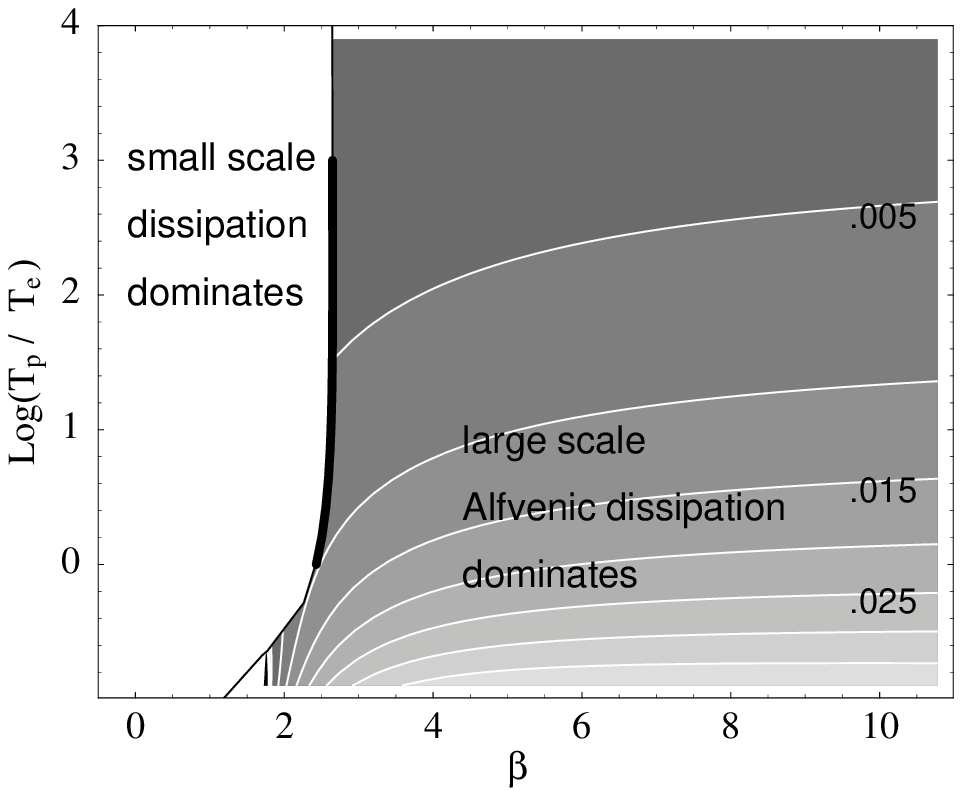}{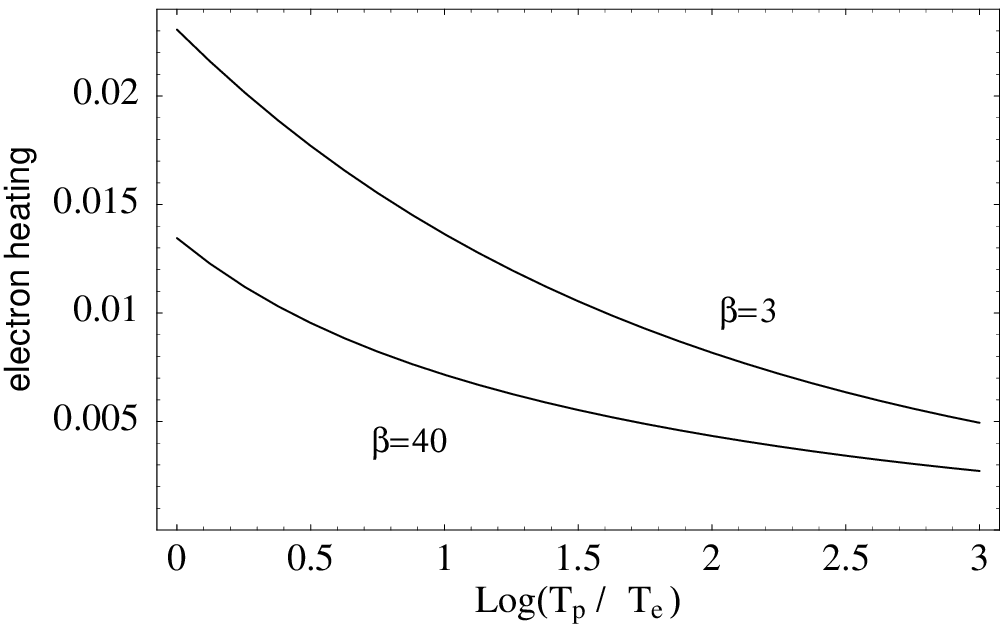}~ \\ ~\\
Figs. \ref{fig-state},\ref{fig-heat}\vskip1cm
\plottwo{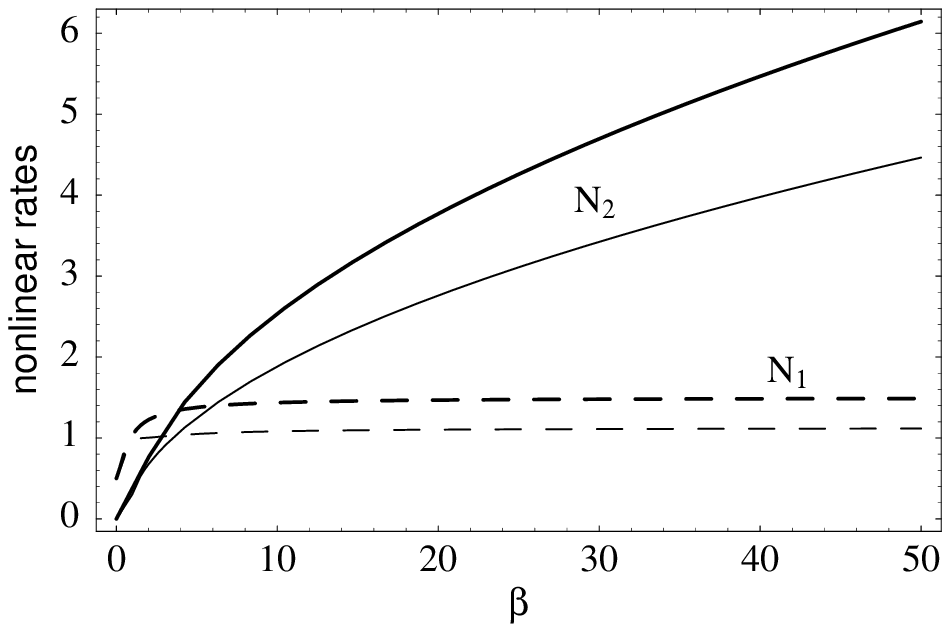}{empty.eps}~ \\ ~\\
Fig. \ref{fig-rates}


\end{document}